\documentclass[preprint, numbers]
{deepmark}
\usepackage{amsmath,graphicx}
\usepackage{amssymb,amsfonts}
\usepackage{algorithm, algpseudocode}
\usepackage{booktabs} 
\usepackage{url}
\title{AWARE: Audio Watermarking via Adversarial Resistance to Edits}

\author{Kosta Pavlović\dmkcorrauthormark}
\author{Lazar Stanarević}
\author{Petar Nedić}
\author{Elena Nešović}
\author{Slavko Kovačević}
\author{Igor Djurović}

\affil{DeepMark}    

\dmkcorrespondence{Kosta Pavlović <kosta@deepmark.me>}
\dmkcode[Code is available at:]{https://github.com/deepmarkpy/aware}
\dmkkeywords{audio watermarking; adversarial perturbations; desynchronization robustness; bitwise readout head}

\begin{document}

\maketitle
\thispagestyle{firststyle}

\begin{abstract}
Prevailing practice in learning-based audio watermarking is to pursue robustness by expanding the set of simulated distortions during training. However, such surrogates are narrow and prone to overfitting. This paper presents AWARE (\textbf{A}udio \textbf{W}atermarking via \textbf{A}dversarial \textbf{R}esistance to \textbf{E}dits), an alternative approach that avoids reliance on attack-simulation stacks and handcrafted differentiable distortions. Embedding is obtained through adversarial optimization in the time–frequency domain under a level-proportional perceptual budget. Detection employs a time–order–agnostic detector with a Bitwise Readout Head (BRH) that aggregates temporal evidence into one score per watermark bit, enabling reliable watermark decoding even under desynchronization and temporal cuts. Empirically, AWARE attains high audio quality and speech intelligibility (PESQ/STOI) and consistently low BER across various audio edits, often surpassing representative state-of-the-art learning-based systems.
\end{abstract}

\section{Introduction}
\label{sec:intro}

Digital watermarking experienced its first major wave of research activity in the 1990s alongside the rapid proliferation of the Internet. Early systems were primarily designed for copyright protection and digital rights management (DRM), with the seminal work of Cox \emph{et al.} introducing spread-spectrum principles to watermarking and setting the agenda for robustness-focused design \cite{cox_spread_spectrum}. Subsequent developments broadened the methodological toolbox with techniques such as quantization index modulation (QIM) \cite{qim} and patchwork-style techniques \cite{patchwork}. While initial approaches were conceived as modality-agnostic and applicable across multimedia, the field soon bifurcated into image- and audio-specific lines of work \cite{img_wm_survey, audio_wm_survey}, each exploiting modality characteristics to improve embedding efficiency and detection reliability. The dominant use case remained copyright protection, driven by the rise of large-scale online content distribution and associated piracy.

Despite their relevance, traditional audio watermarking techniques still exhibit persistent limitations. Beyond insufficient robustness to classical signal processing operations, two challenges are particularly prominent: \emph{desynchronization} and \emph{waveform cuts}. Systems often include dedicated synchronization codes to address time-scale modifications, resampling drift, cropping, and jitter. However, reliably detecting synchronization markers is nearly as difficult as extracting the watermark itself, and thus inherits similar failure modes under distortion. Moreover, many legacy designs embed watermark bits within a single frame or a narrow group of frames, yielding limited temporal redundancy and weak fragment-level detectability. Missing or re-ordered frames, or partial content removal, can therefore break the decoding process and make the watermark hard or impossible to reconstruct.

The advent of modern generative AI has precipitated a renaissance in digital audio watermarking. High-fidelity synthesis models such as: GANs \cite{bigvgan} and diffusion models \cite{audioldm} enable convincing audio and audiovisual “deepfakes” at scale. The risks span reputational harm, fraud, misinformation, and weakened evidence reliability. In response, watermarking has re-emerged as a practical mechanism to label both synthetic and authentic content to support provenance, traceability, and downstream moderation. Accordingly, policy frameworks increasingly cite watermarking among key techniques for AI transparency and content provenance \cite{eu_ai_act}.

Concurrently, the community has begun to “fight fire with fire,” developing end-to-end deep learning (DL) audio watermarking systems. These methods have substantially improved robustness to many standard signal processing distortions and alleviated several long-standing limitations of classical watermarking. Leading this line of research, RobustDNN \cite{robustdnn} defined the basic blueprint, after which WavMark \cite{wavmark} and AudioSeal \cite{audioseal} introduced meaningful improvements. Nevertheless, contemporary benchmark studies still indicate unresolved limitations, with a clear room for progress \cite{benchmark, audiomarkbench}. Watermark decoding for audio under temporal cuts and splicing remains underexplored, aside from zero-bit approaches \cite{desynhDNN} that have limited practical scope. The effects of transmission over the Web are also insufficiently studied, despite their practical importance, as such transmission typically introduces a compound set of distortions whose joint impact on watermark detection can be more severe than that of the individual degradations alone \cite{robustdnn}.



Recent learning-based watermarking methods typically pursue robustness by training against a set of simulated distortions. While effective for the specific attacks seen during training, this strategy suffers from two fundamental limitations. First, the space of plausible audio edits is effectively unbounded, making exhaustive simulation infeasible. Second, models tend to overfit to the chosen distortion set and generalize poorly to unseen edits, especially when real-world degradations occur in combination and produce compound effects that are more harmful than those of the individual attacks considered separately.

In this work, we argue that robustness to such edits should primarily arise from architectural design choices rather than from increasingly complex attack-simulation pipelines. Specifically, audio watermark detection must aggregate weak, distributed evidence over time in a manner that is invariant to ordering, alignment, and signal length. Motivated by this perspective, we introduce AWARE (\textbf{A}udio \textbf{W}atermarking via \textbf{A}dversarial \textbf{R}esistance to \textbf{E}dits), a watermarking framework that combines adversarial embedding under explicit perceptual constraints with a detector architecture explicitly designed for robust audio watermark detection.

\section{Threat model}
We consider a practical audio watermarking threat model in which an adversary aims to disrupt watermark detection while keeping edits imperceptible or near-imperceptible to human listeners. Evaluations are restricted to black-box attacks, assuming no access to detector internals and allowing only signal-level manipulations. This setting reflects realistic deployments, where watermarking systems operate as closed modules and internal details are unavailable to attackers. Accordingly, we target common post-processing and distribution edits, including filtering, lossy compression, additive noise corruption, temporal cuts/cropping, desynchronization (e.g., resampling or time-scale modification), acoustic replay, networked transmission effects (e.g., codec distortion, delay, jitter, and packet loss). We also treat montage/splicing manipulations as in-scope, where content may be cut, removed, or re-ordered. Furthermore, we consider passage through speech-generation and transformation pipelines, including neural vocoder re-synthesis and neural audio compression.

Speech enhancement/separation \cite{sepformer} and voice cloning \cite{valle} are also increasingly relevant. However, these pipelines introduce substantial signal and even semantic alterations. Robustness under such transformations remains an important open problem, as no established solution currently exists in the audio watermarking literature.

\section{Background}

In the standard adversarial setting, we start with a clean input $x$, label $y$, a model $f$, and a task-specific loss $\mathcal{L}$. The objective is to find a norm-bounded perturbation $\Delta$ that maximizes the loss:
$ \max_{\lVert \Delta \rVert_p \le \epsilon}\ \mathcal{L}\big(f(x+\Delta),\, y\big)$. The choice of $p$ (e.g., $p=\infty$ or $2$) and $\varepsilon$ defines the perturbation “budget”.

Adversarial perturbations have proved effective across diverse tasks within the audio modality \cite{commander_song, attack_asr, antifake}, predominantly as tools to break models rather than to enforce desired behaviors. Recent work has explored adversarial formulations of watermark embedding, where watermark is treated as a signal perturbation, optimized to steer the detector’s outputs. Li \emph{et al.} \cite{li_shallow_wm} propose such watermarking framework for images. While the underlying idea of detector-driven adversarial embedding highlights useful principles for audio watermarking, it does not transfer directly to audio signals. In particular, audio watermark detection must contend with variable-length inputs, temporal re-ordering, and partial deletions, which fundamentally differ from the spatial cropping in images.

In our method, we adopt adversarial formulation for audio watermarking and avoid explicit simulation of attacks during optimization. Robustness is instead promoted through the design of a detector architecture tailored to the requirements of audio watermark detection, while watermark embedding is constrained by an explicit, fixed perceptual budget to control signal quality. These choices allow robustness to arise from architectural and optimization constraints rather than from exposure to specific simulated transformations.
\section{AWARE: Method}
\label{sec:method}

\textbf{Watermark Embedding.} Embedding is carried out in the time-frequency (TF) domain, a standard setting for audio watermarking. We modify magnitudes under perceptual constraints and preserve the original phase for reconstruction (iSTFT). The complete procedure is outlined in Algorithm \ref{alg:aware}.

\begin{algorithm}
\caption{AWARE Embedding Procedure}
\label{alg:aware}
\begin{algorithmic}[1]
\Require waveform $x$, watermark $\tilde{w}\in\{-1,+1\}^N$, detector $D$, embedding band $\mathcal{F}=[f_\ell,f_h]$, bin budgets $B$, margin weight $\lambda$, iterations $K$.
\State $M \gets \lvert\mathrm{STFT}(x)\rvert$ 
\State $\Delta \gets 0$ with $\mathrm{supp} (\Delta)\subseteq\mathcal{F}$ \Comment{initialize $\Delta$}
\For{$k=1$ to $K$}
  \State $y \gets D(M+\Delta)$
  \State $\mathcal{L} \gets \frac{1}{N}\|y-\tilde{w}\|_2^2 - \lambda \frac{1}{N}\sum_i |y_i|$ \Comment{push loss}
  \State $\Delta \gets \mathrm{OptimizerStep}(\Delta, \nabla_\Delta \mathcal{L})$
  \State \textbf{for all} $(f,u)$: $\Delta_{f,u} \gets \mathrm{clip}(\Delta_{f,u}, -B_{f,u}, +B_{f,u})$
\EndFor
\State $M' \gets M+\Delta$
\State $\tilde{x} \gets \mathrm{iSTFT}(M', \angle \mathrm{STFT}(x))$ \Comment{reuse original phase}
\Ensure watermarked audio $\tilde{x}$
\end{algorithmic}
\end{algorithm}

Let $x \in \mathbb{R}^T$ be a waveform and let $\lvert \mathrm{STFT}(x)\rvert\in\mathbb{R}_{\ge 0}^{F \times U}$ denote its short-time Fourier transform (STFT) magnitude, with frequency bins $f\in[0,\dots,F-1]$ and time frames $u\in[0,\dots,U-1]$. We restrict perturbations to an audible midband $\mathcal{F}=\{f : f_\ell \le f \le f_h\}$ with $f_\ell=1000~\text{Hz}$ and $f_h=4000~\text{Hz}$ to avoid removal by low/high-pass filters. 

Phase-domain watermark embedding is avoided due to human hearing being largely insensitive to changes in phase, which allows for low-audibility phase manipulations that effectively erase the mark.

We denote $w$ as a watermark of $N$ bits encoded antipodally as $\tilde{w}\in\{-1,+1\}^{N}$. This centers targets at zero, yields balanced gradients, and pairs naturally with margin-based objectives and sign decoding, unlike $\{0,1\}$ coding which pushes scores towards probability bounds and can hinder optimization. In our embedding procedure, the detector $D$ is randomly initialized and its weights are kept frozen throughout optimization. $D$ maps a TF magnitude representation of $x$ to $\left(-1,+1\right)^{N}$ and that provides gradients for the adversarial embedding process.

The embedding procedure minimizes a \emph{push loss} objective that drives the detector toward accurate and confident bipolar decisions on the target bits (increasing the margin to $\pm 1$). Let $M=\lvert \mathrm{STFT}(x)\rvert$ be the STFT magnitude, and let $\Delta$ be a magnitude perturbation supported on $\mathcal{F}$. The detector prediction is $y = D(M+\Delta)\in\left(-1, +1\right)^N$. Optimization objective is given by: 
\begin{equation}
\mathcal{L}_{\text{push}}(M,\Delta; \tilde{w})
\;=\;
\underbrace{\frac{1}{N}\|y - \tilde{w}\|_2^2}_{\text{MSE to targets}}
\;-\;
\lambda \,\underbrace{\frac{1}{N}\sum_{i=1}^{N} |y_i|}_{\text{margin term}},
\label{eq:push-loss}
\end{equation}
with $\lambda>0$ controlling the margin strength.

Rather than imposing a single global norm budget on $\Delta$, we use a per-bin, level-proportional budget that allows larger changes where the signal is louder and smaller changes in quiet regions, consistent with basic psychoacoustics. Let $\tau_{\text{dB}}>0$ be a tolerance parameter (in dB). Its linear amplitude factor is $\eta \;=\; 10^{-\tau_{\text{dB}}/20}.$ For each TF bin $(f,u)$, the admissible magnitude change is bounded by $|\Delta_{f,u}| \;\le\; \eta\, M_{f,u}$. We enforce these bounds via projection (clipping) after each optimizer step. Unlike loss-only quality terms (as in \cite{li_shallow_wm}), this explicit budget provides stronger and more reliable quality control and avoids re-embedding caused by overly aggressive updates. Bin-wise perceptual budgeting is common in classical, masking-inspired watermarking, but it has been largely de-emphasized in recent learning-based systems, where perceptual control is typically folded into a soft term in the overall loss that offers weaker guarantees than the hard constraints. 

In future work the tolerance $\tau_{\text{dB}}$ can be replaced by frequency-dependent thresholds or full psychoacoustic models (threshold-in-quiet and simultaneous masking), enabling budgets that also consider neighboring-frequency masking.

\begin{figure}[tb]
\begin{minipage}[b]{1.0\linewidth}
  \centering
  \centerline{\includegraphics[width=12cm]{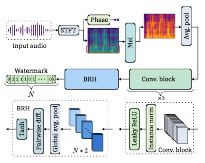}}
\end{minipage}
\caption{AWARE detector architecture.}
\label{fig:arch}
\end{figure}

\noindent \textbf{Watermark detection.} The detector (Fig.~\ref{fig:arch}) is designed to keep activations across all layers stable under realistic audio edits. This leads to detection performance that is less dependent on the specific edit applied. Architectural depth and complexity are deliberately constrained. We argue that constraining architectural expressiveness improves robustness by discouraging reliance on fragile, edit-specific cues and by promoting aggregation of stable evidence. In particular, components that summarize information along the temporal axis reduce sensitivity to misalignment, partial content removal, and other realistic audio edits.


The first layer of the detector computes a Mel spectrogram from the STFT magnitude. The Mel domain aggregates spectral energy into perceptually motivated bands, yielding a representation that is more robust than raw spectrograms to mild time–frequency distortions. Moreover, Mel features are standard in TTS/vocoder pipelines, increasing the chance that a watermark detectable in Mel-space survives voice cloning.

Following ablation findings in \cite{li_shallow_wm}, we include a single pooling layer. This coarsens temporal resolution and improves robustness to local jitter and minor desynchronization.

The next stage contains three feature extraction blocks that intentionally avoid temporal mixing by treating Mel bands as channels and applying 1D convolutions with kernel size $1$ along the temporal axis (stride $1$), followed by instance normalization and a Leaky ReLU. This processes each frame independently, so the filters act as channel (frequency) mixers at a fixed time and avoid temporal mixing.

Instead of using fully connected (FC) layers which bind decisions to absolute positions and fixed input lengths and  become brittle under the effects that alter the index–time mapping, we introduce a Bitwise Readout Head (BRH) that reads out $N$ bits using paired filters. These filters aggregate evidence over time, and produce one position-agnostic score per bit. Concretely, the BRH applies two filter banks to the extracted features $Z\in\mathbb{R}^{C\times U'}$:
\begin{equation}
\begin{aligned}
A^{(0)} &= W^{(0)} Z \in \mathbb{R}^{N\times U'}, \\
A^{(1)} &= W^{(1)} Z \in \mathbb{R}^{N\times U'}
\end{aligned}
\end{equation}

\noindent with $W^{(0)},W^{(1)}\in\mathbb{R}^{N\times C}$, so that for each bit index $i$ we obtain two activation traces $A^{(0)}_{i,\cdot}$ and $A^{(1)}_{i,\cdot}$ along time. The traces are then aggregated by global averaging:
\begin{equation}    
\bar a^{(b)}_i \;=\; \frac{1}{U'} \sum_{u=1}^{U'} A^{(b)}_{i,u}, \quad b\in\{0,1\},
\end{equation}
and contrasted to produce a single bit score:
\begin{equation}
    g_i \;=\; \bar a^{(1)}_i - \bar a^{(0)}_i, \qquad y_i \;=\; \tanh(g_i).
\end{equation}
Here, $\tanh(\cdot)$ is a monotone squashing function that maps scores to $(-1,+1)$. The output $y=(y_1,\ldots,y_N)$ thus provides one position-agnostic score per watermark bit, obtained by temporal evidence aggregation within the BRH.

Having described the mechanics of the BRH, we now sketch the intuition behind its structure. Convolutional filters are typically crafted to fire on specific stimuli or patterns (e.g., in image classification/object detection, some filters respond to “ears”, others to “eyes”, etc.). In watermark detection, those stimuli are bits. Accordingly, the BRH allocates two filters per bit $b$: one tuned to evidence for $b=1$ and one for $b=0$. Each filter produces a temporal activation trace that global averaging converts into evidence scores. The bit decision becomes a simple competition. Whichever filter accumulates more evidence over time, “wins”.

In addition to bit extraction, the BRH can be used to determine whether a given audio snippet is watermarked. Let $y = \mathrm{BRH}(x)$ denote the vector of per-bit scores produced by the detector. We define the confidence of the watermark decision as $\mathrm{conf}(x) \;=\; \frac{1}{N} \sum_{i=1}^{N} |y_i|$, i.e., the mean absolute activation of all BRH outputs. Intuitively, this score measures the overall strength of the aggregated watermark evidence. A detection threshold $\tau_{\text{conf}}$ is applied to determine watermark presence. If $\mathrm{conf}(x) > \tau_{\text{conf}}$, the signal is considered watermarked; otherwise, it is treated as non-watermarked. This is particularly useful under splicing and cutting, as BRH activations peak when the detector is aligned with segments that contain the watermark. Consequently, the confidence score can be evaluated over sliding or segmented windows and localize watermarked regions within a longer audio.

\section{Experimental Setup}
We compare against the strongest publicly available baselines (WavMark and AudioSeal) on $2000$ approximately $5$-second samples from the LibriSpeech \cite{librispeech} dataset at a $16$\,kHz sampling rate. Perceptual quality and speech intelligibility are evaluated using PESQ \cite{pesq} and STOI \cite{stoi}, while watermark robustness is measured by bit error rate (BER). We evaluate message-level watermark detection using the false negative rate (FNR). The detection threshold $\tau_{\text{conf}}$ is set such that $\text{FPR} < 0.01$ under no perturbations ($0.04$ in our setting).

For a fair comparison, all methods are tested with payloads of $16$\,bps, matching the configuration of the comparative baselines. Detection thresholds are calibrated to achieve a comparable $\text{FPR} < 0.01$ on clean, held-out audio. We note that the original AudioSeal evaluation tunes thresholds separately for each attack. We instead use a fixed threshold of $0.5$ across all conditions, as per-attack thresholding assumes prior knowledge of distortions and may not reflect realistic deployment settings.

In addition, we report results for AWARE at $20$\,bps capacity by partitioning the signal into $1$\,s segments and embedding $20$ bits per segment. Detection is performed by sliding-window with a fixed hop of $40$\,ms and retaining windows whose BRH confidence exceeds a specified threshold $\tau_{\text{conf}}$, after which the final message is obtained by bitwise majority voting across detections. Segment length is adjusted to the smallest multiple of the hop size that is greater than or equal to the original length. This ensures full frame coverage in the STFT and prevents sample loss during the iSTFT reconstruction. For this higher-capacity setting, $\tau_{\text{conf}}$ is increased to $0.06$ to maintain a comparable false-positive rate.

Adversarial embedding is optimized for $K=500$ iterations using the NAdam optimizer \cite{nadam} (learning rate $0.1$), with a reduce-on-plateau scheduler (factor $0.9$). The push-loss margin weight is set to $\lambda=0.1$. STFTs are computed with a frame length of $1024$, hop length $256$, and a Hann window, with embedding restricted to $[1000,\,4000]$~Hz and detection using $128$ Mel bands. After optimization, the signal is reconstructed via iSTFT before detection.

Robustness is probed under: low/high-pass filtering (LPF at $4$\,kHz, HPF at $1000$\,Hz), PCM quantization to $8$\,bits, MP3 compression at $64$\,kbps, Gaussian and pink noise (GN; $20$\,dB, PN; peak at $0.03$), resampling (RS) to $32$\,kHz, and time–scale modification (TS; $\pm 35\%$). We evaluate resistance to band-stop (notch) filtering (BSF) with a $200$\,Hz-wide notch. The notch center is swept across the $0$–$4000$\,Hz range in $100$\,Hz steps, and we report the worst-case error rate over all positions. Recent work reports vulnerabilities to such notches in models like AudioSeal \cite{bas}. We test pitch shift (PS; $10$ cents), vocoder resynthesis (NV) using BigVGAN \cite{bigvgan}, and neural audio compression (NAC) with EnCodec \cite{encodec}, for which benchmarks \cite{benchmark, audiomarkbench} indicate limited robustness across many systems.

To explicitly evaluate montage and splicing scenarios, we consider two attacks. Segment removal (SR) removes a single contiguous segment corresponding to $50\%$ of the audio. Permutation attack (PA) simulates re-ordering by splitting the signal at a random point and swapping the segments.

Beyond signal-level edits, we consider two attacks representative of real-world scenarios. These attacks capture compound effects that arise in practical deployments and provide a realistic stress test of watermark robustness. First, replay attack (RA) \cite{replay_attack} simulates re-recording through acoustic environments using impulse responses from the Aachen Impulse Response database \cite{air_db}, combined with mild Gaussian noise ($50$\,dB) and band-pass filtering ($50$\,Hz - $8$\,kHz) to reflect microphone and loudspeaker characteristics. Second, a VoIP transmission attack (VT) emulates WebRTC-based pipelines \cite{webrtc} with Opus encoding \cite{opus} ($16$\,kbps, $20$\,ms frames) and network impairments (delay=$50$\,ms, jitter=$20$\,ms, loss=$2\%$).

\begin{figure}[t]
  \centering
  \includegraphics[width=8.5cm]{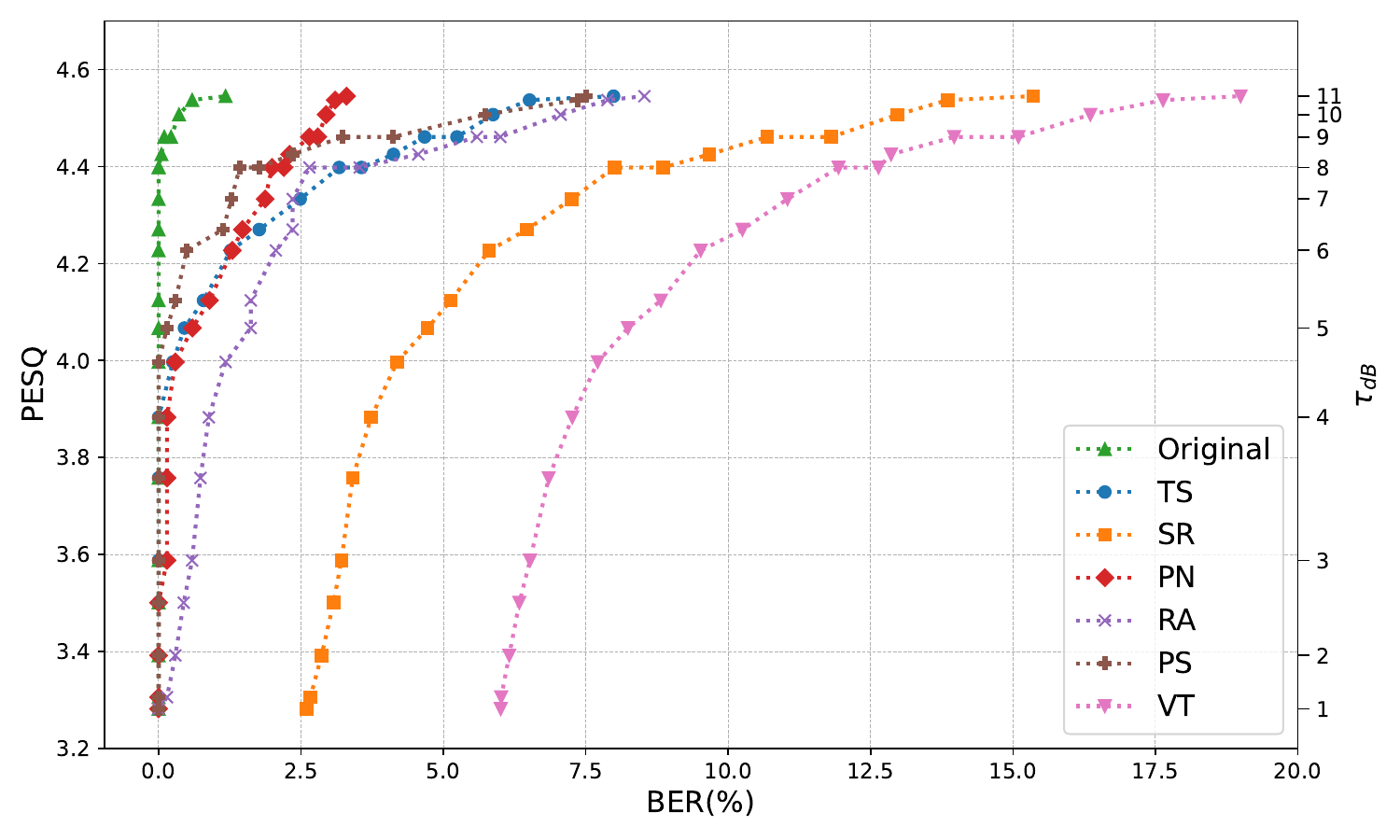}
  \caption{Quality--robustness trade-off controlled by the perceptual budget $\tau_{\text{dB}}$.}
  \label{fig:quality_robustness}
\end{figure}

\begin{table}[b]
\centering
\small
\caption{Speech quality and intelligibility (mean $\pm$ std).}
\label{tab:quality}
\begin{tabular}{l|c|c}
\toprule
\textbf{Method} & \textbf{PESQ} $\uparrow$ & \textbf{STOI} $\uparrow$ \\
\hline
WavMark   & $4.23 \pm 0.18$ & $0.99 \pm 0.009$ \\
AudioSeal & $4.43 \pm 0.08$ & $0.99 \pm 0.002$ \\
AWARE     & $4.26 \pm 0.09$ & $0.99 \pm 0.003$ \\
AWARE (20bps) & $4.24 \pm 0.12$ & $0.99 \pm 0.004$ \\
\bottomrule
\end{tabular}
\end{table}

\section{Results and Analysis}

The quality--robustness trade-off (BER vs. PESQ) across distortions is shown in Fig.~\ref{fig:quality_robustness}. The proposed model consistently operates in the upper-left region, corresponding to low BER and high perceptual quality. This indicates that the robustness comes with minimal perceptual cost and provides a practical region for parameter selection. Based on this trade-off, we select $\tau_{\mathrm{dB}}=6.2$ as the operating point for the remaining experiments reported in Tables~\ref{tab:quality} and~\ref{tab:ber_attacks}. Table~\ref{tab:quality} shows that all methods achieve similarly high audio quality and speech intelligibility. Overall, the watermark remains \emph{imperceptible} or \emph{near-imperceptible} across systems.

Robustness results in Table~\ref{tab:ber_attacks} demonstrate that AWARE remains effective across diverse edits, with only small fluctuations in BER between conditions. AudioSeal degrades sharply on spectral edits (LPF, HPF and BSF), while WavMark fails under PCM and PS. Under noise corruption (GN and PN), AWARE outperforms comparative methods, indicating strong resilience to background noise and suggesting that the AWARE watermark is less noise-like and more structurally integrated into the signal. Passage through a NV yields very low error rate for AWARE, whereas baselines struggle. AudioSeal outperforms AWARE for NAC, likely due to its explicit exposure to such distortions during training. RS does not significantly impact any of the evaluated systems. Under TS, AWARE attains the lowest BER among all evaluated methods. SR proves to be a challenging case, as AWARE incurs higher BER, though still within a usable range and comparable to the strongest baseline. PA has little effect on AWARE, which confirms that time-order-agnostic aggregation mitigates sensitivity to reordering. 

\begin{table}
\centering
\footnotesize
\setlength{\tabcolsep}{3pt}
\renewcommand{\arraystretch}{1.05}
\caption{BER with message-level FPR/FNR in parentheses.}
\label{tab:ber_attacks}
\begin{tabular}{l|c|c|c|c}
\toprule

Cond. & WavMark & AudioSeal & AWARE  & \shortstack{AWARE\\(20\,bps)} \\
\midrule
Original & \textbf{0.000} {\tiny(0.00/0.00)} & \textbf{0.000} {\tiny(0.00/0.00)} & \textbf{0.000} {\tiny(0.00/0.00)} & \textbf{0.000} {\tiny(0.00/0.00)} \\
LPF      & \textbf{0.000} {\tiny(0.00/0.00)} & 0.146 {\tiny(0.00/0.00)} & \textbf{0.000} {\tiny(0.00/0.00)} & \textbf{0.000} {\tiny(0.00/0.00)} \\
HPF      & \textbf{0.000} {\tiny(0.00/0.00)} & 0.409 {\tiny(0.00/0.00)} & \textbf{0.000} {\tiny(0.00/0.00)} & \textbf{0.000} {\tiny(0.00/0.00)} \\
BSF      & \textbf{0.000} {\tiny(0.00/0.00)} & 0.338 {\tiny(0.00/1.00)} & 0.010 {\tiny(0.00/0.00)}  & \textbf{0.000} {\tiny(0.00/0.01)} \\
PCM      & 0.100 {\tiny(0.00/0.11)} & \textbf{0.005} {\tiny(0.00/0.01)} & \textbf{0.000} {\tiny(0.00/0.00)} & \textbf{0.000} {\tiny(0.00/0.00)} \\
MP3      & 0.023 {\tiny(0.00/0.02)} & \textbf{0.000} {\tiny(0.00/0.00)} & \textbf{0.001} {\tiny(0.00/0.00)} & \textbf{0.003} {\tiny(0.00/0.00)} \\
PS       & 0.500 {\tiny(0.00/0.75)} & 0.100 {\tiny(0.00/0.28)} & \textbf{0.014} {\tiny(0.00/0.00)} & \textbf{0.004} {\tiny(0.00/0.00)} \\
GN       & 0.460 {\tiny(0.00/0.51)} & 0.033 {\tiny(0.00/0.06)} & \textbf{0.012}  {\tiny(0.00/0.01)} & \textbf{0.000} {\tiny(0.00/0.14)} \\
PN       & 0.286 {\tiny(0.00/0.75)} & 0.087 {\tiny(0.00/0.68)} & \textbf{0.015} {\tiny(0.00/0.02)} & \textbf{0.012} {\tiny(0.00/0.09)} \\
NV       & 0.500 {\tiny(0.00/1.00)} & 0.390 {\tiny(0.00/1.00)} & \textbf{0.016} {\tiny(0.00/0.00)} & \textbf{0.009} {\tiny(0.00/0.00)} \\
NAC       & 0.500 {\tiny(0.00/1.00)} & \textbf{0.091} {\tiny(0.00/0.07)} & 0.179 {\tiny(0.00/0.70)} & 0.260 {\tiny(0.00/0.98)} \\
RS       & \textbf{0.000} {\tiny(0.00/0.00)} & \textbf{0.000} {\tiny(0.00/0.00)} & \textbf{0.000} {\tiny(0.00/0.00)} & \textbf{0.000} {\tiny(0.00/0.00)} \\
TS       & 0.257 {\tiny(0.00/0.19)} & 0.068 {\tiny(0.00/0.93)} & \textbf{0.016} {\tiny(0.00/0.00)} & \textbf{0.027} {\tiny(0.00/0.40)} \\
SR       & 0.037 {\tiny(0.00/0.07)} & \textbf{0.007} {\tiny(0.00/0.21)} & 0.063 {\tiny(0.00/0.10)} & 0.021 {\tiny(0.00/0.04)} \\
PA       & \textbf{0.000} {\tiny(0.00/0.00)} & 0.042 {\tiny(0.00/0.37)} & \textbf{0.000} {\tiny(0.00/0.00)} & \textbf{0.000} {\tiny(0.00/0.00)} \\
RA       & 0.500 {\tiny(0.00/1.00)} & 0.500 {\tiny(0.00/1.00)} & \textbf{0.024} {\tiny(0.00/0.11)} & \textbf{0.029} {\tiny(0.00/0.58)} \\
VT       & 0.239 {\tiny(0.00/0.44)} & 0.480 {\tiny(0.00/1.00)} & \textbf{0.102} {\tiny(0.00/0.23)} & \textbf{0.104} {\tiny(0.00/0.16)} \\
\bottomrule
\end{tabular}
\end{table}

The two compound, real-world attacks further highlight the benefits of the proposed design. Under RA, AWARE retains reliable detection, whereas both baselines fail. Similarly, under VT, AWARE significantly outperforms comparative methods. These results indicate that robustness achieved through architectural design generalizes beyond isolated signal-level edits to more realistic deployment scenarios. 

AWARE ($20$\,bps) follows similar trends and can outperform utterance-level embedding in several conditions (BSF, GN, PN, NV, and SR) due to sliding-window inference with majority voting, which introduces temporal redundancy and increase the chance of high-confidence detections. For scenarios with higher message-level FNR (NAC, TS, and RA), the number of available windows appears insufficient given the short LibriSpeech utterances ($\approx 5$\,s). Longer signals should enable more reliable aggregation and improved detection. Overall, the results suggest that AWARE scales well with increased payload.

We again note that the results for the AWARE system are obtained \textbf{\textit{without}} direct simulation of specific attacks during training. Robustness largely stems from the detector architecture and the BRH. Conversely, competing systems often see (and thus favor) certain distortions during training, which helps in those particular cases but can leave gaps elsewhere. 

\section{Ablation Studies}
To evaluate effectiveness and rationale of key architectural and representational design choices, we
conduct ablation studies. Each experiment isolates one architectural or representational component while keeping all other settings fixed. This allows us to examine how each component impacts robustness and consistency under typical audio distortions. For every ablation, we chose representative attack subsets that the ablated component is expected to affect. The results are summarized in the tables below as mean BER across those cases.

\subsection{BRH vs. Fully Connected Output}

Table~\ref{tab:brh} ablates the BRH, as a key architetural component, against a conventional FC output layer, preceded by global pooling over the temporal dimension. BRH markedly improves performance on edits that disrupt alignment or ordering (TS, PA, RA and VT), as it aggregates evidence over time without relying on absolute positions. By contrast, the FC variant collapses once the input continuity is broken.

\begin{table}
\centering
\footnotesize
\setlength{\tabcolsep}{3.5pt}
\renewcommand{\arraystretch}{1.05}
\caption{BER with and without BRH.}
\label{tab:brh}
\begin{tabular}{l|*{7}{c}}
\toprule
\textbf{Model} & \textbf{PS} & \textbf{NAC} & \textbf{TS} & \textbf{SR} & \textbf{PA} & \textbf{RA} & \textbf{VT} \\
\midrule
w/ BRH  & \textbf{0.008} & \textbf{0.179} & \textbf{0.016} & \textbf{0.095} & \textbf{0.000} & \textbf{0.022} & \textbf{0.095} \\
w/o BRH & 0.359 & 0.470 & 0.373 & 0.170 & 0.386 & 0.404 & 0.170 \\
\bottomrule
\end{tabular}
\end{table}

\subsection{With vs. Without Mel Projection}
Finally, we compare the proposed Mel-based detector with a variant operating directly on STFT magnitudes. The Mel front-end performs perceptual band aggregation, which stabilizes frequency-domain statistics and aligns the representation with human-critical bands and common synthesis pipelines. Accordingly, we probe attacks that directly stress these properties: LPF/HPF, MP3, NV, RA and VP, precisely where Mel aggregation is expected to enhance the robustness.

\begin{table}[h]
\centering
\footnotesize
\setlength{\tabcolsep}{3.5pt}
\renewcommand{\arraystretch}{1.05}
\caption{BER with and without the Mel front-end.}
\label{tab:mel}
\begin{tabular}{l|*{7}{c}}
\toprule
\textbf{Front-End} & \textbf{LPF} & \textbf{HPF} & \textbf{MP3}  & \textbf{PS} & \textbf{NV} & \textbf{RA} & \textbf{VT}  \\
\midrule
w/ Mel (proposed) & \textbf{0.000} & \textbf{0.000} & \textbf{0.001} &  \textbf{0.014}  & \textbf{0.016} & \textbf{0.024} & \textbf{0.102}  \\
w/o Mel (STFT only) & 0.007 & 0.008 & 0.014 & 0.147 &  0.530 &  0.058 &  0.136  \\
\bottomrule
\end{tabular}
\end{table}

Results in Table \ref{tab:mel} suggest that for spectral edits (LPF/HPF) and compression (MP3), absolute differences are modest, yet consistently favor the Mel front-end. Under PS and NV resynthesis, the gap becomes substantial. The Mel-based detector remains highly reliable, whereas the STFT-only variant degrades sharply. A clear, though more moderate, advantage is also observed under RA, which indicates that Mel aggregation provides additional robustness to acoustic channel coloration and recording/playback artifacts. Under VT, the Mel front-end again improves performance, and shows better tolerance to the compound distortions introduced by networked speech communication pipelines.




\section{Conclusion}
This paper advocates \emph{robustness by design} for digital audio watermarking and introduces a time--order--agnostic detector with a Bitwise Readout Head. The proposed system achieves high audio quality and low error rate across a wide range of audio edits, often surpassing strong learning-based baselines despite not being trained on attack-specific distortions. However, rather than pursuing “state-of-the-art” claims, particularly tenuous in this domain, given competing metrics and trade-offs, we aim to provide principled guidance on design choices aligned with practical threat models for digital audio watermarking use cases.

\bibliographystyle{unsrtnat} 

\end{document}